\documentclass[sigconf]{acmart}

\usepackage{booktabs} % For formal tables
\usepackage{subfigure}

\setcopyright{rightsretained}
\acmDOI{10.475/123_4}

\acmISBN{123-4567-24-567/08/06}

\acmConference[ACM Digital Health'18]{ACM Digital Health Conference}{April 2018}{Lyon, France} 
\acmYear{2018}
\copyrightyear{2018}

\acmArticle{4}
\acmPrice{15.00}

\begin{document}
% \title{Enriching Demographic Descriptors with    
% Social-Media Traces for Environmental Epidemiology}

% \title{Diagnosing Population Health across Geographies with Social Media Sensors}
% \title{A Data-driven Approach to Predicting Ambulance Calls Across Geographies}
% \title{Uncovering The Geographic Divide in Epidemiology Using Ambulance Calls Data}
% \title{Mobile Web to The Rescue: \\ Predicting Ambulance Calls Across Geographies}
% \title{Location-based Services to The Rescue: \\ Predicting Ambulance Calls Across Geographies}
\title{Foursquare to The Rescue: \\ Predicting Ambulance Calls Across Geographies}

% \author{ExampleAuthor1}
% %\authornote{The secretary disavows any knowledge of this author's actions.}
% \affiliation{%
%   \institution{Institution1}
%   %\streetaddress{P.O. Box 1212}
%   \city{Institution2} 
%   %\state{Ohio} 
%   %\postcode{43017-6221}
% }
% \email{email}

% \author{ExampleAuthor2} 
% \affiliation{%
% 	\institution{Institution}
% 	% \streetaddress{Rono-Hills}
% 	\city{Adress} 
% 	% \state{Arunachal Pradesh}
% 	% \country{India}}
% }
% \email{email}

% \title{Mobile Web to }
 \author{Anastasios Noulas}
 \affiliation{%
   \institution{New York University}
   \city{New York} 
   \state{USA} 
 }
 \email{noulas@nyu.edu}

 \author{Colin Moffatt}
 \affiliation{%
 \institution{De Montfort University}
   % \institution{North West Ambulance Service, NHS}
   \city{Leicester} 
   \state{UK} 
 }
 \email{syrphus7@gmail.com}

  \author{Desislava Hristova}
 \affiliation{%
   \institution{New York University}
   \city{New York} 
   \state{USA} 
 }
 \email{desii.hristova@gmail.com}

 \author{Bruno Gon\c{c}alves}
 \affiliation{%
   \institution{New York University}
   \city{New York} 
   \state{USA} 
 }
 \email{bgoncalves@gmail.com}

\begin{abstract}

Understanding how ambulance incidents are spatially distributed can shed light to the epidemiological dynamics of geographic areas and inform healthcare policy design. 
% In operational terms, being able to predict ambulance incidents
% translates to quicker response times for paramedic crews and increased numbers of lives saved. 
Here we analyze a longitudinal dataset of more than four million ambulance calls across a region of twelve million residents in the North West of England. With the aim to explain geographic variations in ambulance call frequencies, we employ a wide range of data layers including open government datasets describing population demographics and socio-economic characteristics, as well as geographic activity in online services such as Foursquare. Working at a fine level of spatial granularity we demonstrate that daytime population levels and the deprivation status of an area are the most important variables when it comes to predicting the volume of ambulance calls at an area. Foursquare 
check-ins on the other hand complement these government sourced indicators, offering a novel view to population nightlife and commercial activity locally. We demonstrate how check-in activity can provide an edge when predicting certain types of emergency incidents in a multi-variate regression model.
\end{abstract}


\maketitle

\section{introduction}
\label{sec:intro}
Effectively predicting the demand for ambulances across regions can both improve the operational capacity of emergency services as well as reduced costs by optimizing  resource utilization and providing an optimal spatial deployment and duty planning of paramedic crews. This  results in quicker response times in attending emergency incidents reducing fatalities. Moreover, as ambulances play a critical role as first responders, calls to the ambulance service provide precious real time epidemiological information traces that can  assist population health monitoring at scale and lead
to improved policy design in healthcare. 

Past studies aiming to explain geographic variations in the volume of calls for emergencies~\cite{jones2005circadian, carter2001scheduling, o2013system, peacock2006emergency} have been limited to examining epidemiological patterns across very broad geographic scales (national level). Enabling predictions at finer spatial scales, e.g. at the level of city neighborhoods, can generate intelligence that 
will allow the targeting of healthcare interventions in a more accurate manner, specializing treatment to the
characteristics of populations in need. The importance of geography for health 
in fact has been highlighted through works pointing out that postal code may be a better predictor, compared to genetic information,  when it comes to explaining the well being of local populations~\cite{graham2016your, chetty2016association}. 

% Further, in past research literature
% the mobility dynamics of populations have been ignored. 
In this work, our aim is to estimate the volume of ambulance calls at the level of individual Lower Super Output Areas (LSOA) in the North West of England. 
We investigate various environmental, socio-economic and demographic factors that can contribute to the rise of emergency incidents at a given \textit{locale}. In this setting, we note how a population's level of deprivation   
has a deterring impact on its health status with deprived urban areas being 
those where the number requests for emergency medical attention surges. 
Additionally, we identify regional population volume dynamics as one of the primary drivers for emergency calls to take place, showing how health incidents are likely to occur in areas where people become active and not simply those where they are registered as residents according to census. \emph{Critically to the novelty of the present work, we exploit place semantics and mobility patterns in location based-service Foursquare to infer population activity trends at local areas and attain
more accurate predictions of the volume of calls an area will experience.} 
Our research findings are described in more detail next:
\begin{itemize}
\item{\textbf{Ambulance calls concentrate in urban areas and form patterns of spatial co-occurrence:}} in Section~\ref{sec:scope} we show how the spatial distribution  of calls to emergency services is highly skewed, with a large fraction of activity being concentrated in major urban centers. 
Furthermore, there are strong patterns in terms of how incident types spatially co-occur. For instance, \textit{overdose/poisoning} cases are highly correlated across geographies with \textit{convulsions/seizures} and \textit{unconscious/fainting} incidents. On the other hand \textit{breathing problems} tend to correlated more
with \textit{chest pain} complaints and \textit{sick person} cases.

\item{\textbf{Higher regional levels of deprivation imply higher volumes of ambulance calls:}} In Section~\ref{sec:socio} characterize geographic areas using various socio-economic indicators accessed through open government datasets, including we scores of
geographic regions~\cite{noble_measuring_2006}. We find that \textit{breathing problems}, \textit{chest pain} and \textit{psychiatric/suicide} related incidents are more common in areas with higher crime rates and lower income levels.  

\item{\textbf{Daytime population levels are a better predictor of ambulance calls than residential
population:}} In Section~\ref{sec:population} we define a variable to estimate \textit{daytime population} levels. This is the sum of the number of workers at an area, younger (below 16 year old) and older (above 65 year old). We describe the stark differences
between the spatial-distribution of \textit{daytime} and \textit{residential} populations 
showing how the former yields a much higher correlation score (pearson's $r=0.68$ vs $r=0.18$) with the total number of calls in an area and is key to explaining variations for
a set of incidents types including \textit{traumatic/injuries} and \textit{uconscious/fainting}. 

\item{\textbf{Foursquare activity patterns at urban regions contribute 
to better predictions in ambulance calls for an area:}}
Finally, in Section~\ref{sec:results} we formulate a prediction task where our goal  becomes to combine the various information sources discussed above and assess their relative importance in predicting the number of ambulance calls at an area. Daytime population levels are the most significant factor in explaining variations in ambulance calls, followed by the index of multiple deprivation for an area and foursquare check-in activity. The importance of each variable however, depends on the type of incident considered. Daytime population level best explain high number in \textit{falls} and \textit{traumatic injuries} whereas calls at areas with increased levels of \textit{unconscious/fainting} and \textit{overdose/poisoning} incidents are best approximated using check-in frequencies from location-based service Foursquare. The service proves to be a useful proxy of population activity at Food and Nightlife places.    
\end{itemize}
\vspace{-0.08cm}
Our work demonstrates how traditional, yet critical, sectors in healthcare
may be improved through the integration of digital datasets from online sources. Location-based technologies could improve the operational efficiency of emergency services through more refined descriptions of population activities at fine geographic
scales. From an epidemiological perspective they can
be integrated with socio-economic and demographic indicators to offer a deeper understanding on population health patterns. 

\section{Related Work}
\label{sec:related}
\begin{table*}[]
\centering
\begin{tabular}{lll}
\hline
Source & Variable Description  \\
\hline
\textbf{Lower Super Output Area Boundaries (data.gov.uk)}  & LSOA shapefile polygons   \\
\textbf{2011 Census; Table PHP01 (data.gov.uk)}  & \#People residing in LSOA  \\
\textbf{Communal Population} (CmmnlRs)    & \#People in communal establishments  \\
\textbf{Area size}  & LSOA area in hectares  \\
\textbf{Average Household Size} (AvHshlS) & \#persons in household (mean) \\
\textbf{Workplace Population 2011}\footnote{www.nomisweb.co.uk/census/2011/wp101ew} & Employed People 16-74 \\
\textbf{LSOA Mid-Year Population Estimates 2011 (ons.gov.uk)}   & \# Persons at each age in years   \\
\textbf{English Indices of Multiple Deprivation (www.gov.uk)}   & Deprivation scores for a local area                \\
% Income Deprivation     & Area's rank on median income           \\
% \textbf{Healthcare Providers (NWAS)} & various & Healthcare providers by category \\
\textbf{Foursquare check-in data}  & logged check-ins and place categories by LSOA     \\        \hline                     
\end{tabular}
\caption{Summary of external data sets used. In cases where abbreviations have been
used they are put in parentheses in the first column.}
\label{tab:extradata}
\end{table*}

\paragraph{\textbf{Health Geography and environmental epidemiology}}
Spatial analysis and health geography trace their roots back to $1854$ London when John Snow famously drew maps with markers of health incidents to locate the source of a cholera outbreak~\cite{snow1855mode}. 
Spatial epidemiology has ever since contributed to our understanding of how diseases spread and appear geographically~\cite{gatrell1996spatial} tracing its roots on spatial statistics and quantitative geography~\cite{fotheringham2000quantitative}. 
The state-of-the-art in statistical epidemiology of non-infectious disease usually treats occurrences as a spatio-temporal point process\cite{moller1998log, diggle2007model, diggle2005point}. This family of techniques however focuses on predicting incident frequencies of a single disease and does not provide interpretations on the environmental and demographic factors that may drive the spatio-temporal occurrence of medical incidents. They trace their roots in methods such as kriging~\cite{stein2012interpolation} and they effectively reduce the problem of modeling the spatio-temporal occurrences of epidemiological incidents to a form of interpolation. 

More recently, research literature in the field of health geography
has focused on explaining geographic variations in emergency requests in terms of census and demographics data which has become available on a national level~\cite{ong2009geographic}. Deprivation levels of urban communities (e.g. accessibility to employment) or differences between rural and urban areas have been projected to partially explain geographic variations in the volume of calls for emergencies~\cite{o2013system, peacock2006emergency}. Our work considers urban deprivation factors in 
explaining call variations at a fine level of spatial granularity of areas with a few hundred residents. Moreover, we investigate the interplay of environmental, demographic and urban activity factors across different incident types (e.g. psychiatric, assault, fall etc.).
\paragraph{\textbf{Social media in health analytics.}}
The field of digital health has risen in recent years thanks to the proliferation of
the web as well as mobile sensing technologies~\cite{servia2017mobile}. Despite their biases~\cite{lazer2014parable}, web and online social media sources, provide ample opportunity to break away many of the barriers that characterize traditional experimental methodology in medical studies. These include being able to track users health behavior in a social context and anonymously~\cite{de2016discovering, de2014mental}, or at large population
scales~\cite{paul2016social} while retaining the benefits of fine spatio-temporal views 
on user behavior~\cite{mejova2015foodporn}. 
An aspect of novelty in the present work regards the incorporation of information 
from social media services to understand ambulance demand regionally. Geo-referenced datasets from services like Foursquare have the advantage of providing us with place semantics and mobility patterns described at fine spatial scales.

\section{Operational scope \& data}
\label{sec:scope}
\subsection{The North West Ambulance Service}
The North West Ambulance Service NHS Trust (NWAS)\footnote{\url{http://www.nwas.nhs.uk/}} is the second largest ambulance trust in England, providing services to a population of seven million people across a geographical area of approximately 5,400 square miles. The organisation provides a 24 hour, 365 days a year accident and emergency services to those in need of emergency medical treatment and transport, responding to hundreds of thousands emergency calls per year. Highly skilled staff provide life-saving care to patients in the community and take people to hospital or a place of care if needed. Calls that result in an ambulance dispatch may come via the $999$ (The Europe-wide $112$ also 
results in a $999$ call). The call operator will ask the caller a series of questions to ascertain the degree of emergency and will assign a dispatch code to the call. Dispatch codes are numeric and correspond to a broad
classification of incidents (e.g. falls, traumatic injuries, 
assualt, psychiatric etc.). If the call requires a response, an appropriate team receives the instruction, and then swiftly makes its way to the incident location, using an onboard satelite navigation system.

\subsection{Datasets}
We next describe the characteristics of the data employed in the present work. The primary source of data, the ambulance calls, has been collected by the North West Ambulance Service (NWAS). We employ numerous datasets to design a number of variables from demographic, socio-economic and web sources (Foursquare). These
data layers will let us assess the efficacy of various information sources in predicting geographic variations in ambulance calls.
% \\
\paragraph{\textbf{Ambulance calls dataset}}
The data provided by NWAS are those routinely collected as an emergency call operator receives a call where it is anticipated an ambulance may be required. The data is  comprised of $4.4$ million calls the service has responded to from April 2013 to March 2017. Each incident has a \textit{dispatch code} which corresponds to the type of the medical condition or cause that led to the call (e.g. suicide, fall, traffic incident etc). Codes range from 1 to 35 though other codes for rare cases may
also be used.  We exploit the incident number to stratify ambulance calls by nature and ask the question whether different incident types are associated with different factors. Critically to the present work, we exploit geographic information on where the incident took place at the administrative level of \textit{lower super output areas} (LSOAs) which in the UK corresponds to the first four letters of the zip code.
We explain LSOAs in detail next. 

\paragraph{\textbf{Spatial boundaries, populations, demographics and socio-economic indicators}}
Table~\ref{tab:extradata} summarizes the additional data sets in terms of the variables that we utilise.
The Lower Super Output Areas (LSOAs) are the fundamental unit of spatial aggregation considered in this work. Output areas were originally created such that populations were approximately similar socially and in size~\cite{cockings2011maintaining}. LSOAs were assembled to maintain such similarity with a target population size of around 1500, but naturally, there is some variation as we demonstrate in the next section.

In terms of population data, we use information on the number of people residing in each LSOA, the number of people in communal establishments (communal population) and average household size. The workplace population corresponds to an enumeration of the people that work in an LSOA. In Section~\ref{sec:analysis} we combine workplace population with residential population of young and older age groups (non working populations) to define the \textit{daytime population} variable, which becomes one of the best predictors for ambulance calls. 

In terms of socio-economic indicators, we employ \textit{The Index of Multiple Deprivation} (IMD) which is a score calculated by the government in the UK to characterize areas through the consideration of a set of deprivation and quality of life indicators. These include income and crime  levels, accessibility to education, health deprivation and disability, barriers to housing as well as the quality of the living environment. IMD has been shown to be a very important discriminative signal when aiming to predict the dynamics of complex urban processes including gentrification~\cite{hristova2016measuring}. 
The number of \textit{healthcare providers} corresponds to an enumaration of health services
that are present in an area, which includes hospitals, general practioners (GPs) and social support facilities amongst others. 

\paragraph{\textbf{Location-based services.}}
Finally, we employ a dataset of public Foursquare check-ins pushed on Twitter, collected over 11 months in 2011. For every check-in information about the place the user has checked-in becomes 
available, including its location in terms of latitude and longitude coordinates. Additionally,
for every Foursquare venue we know the category of it (Coffee Shop, Italian Restaurant etc.). 
For the purposes of the present work we use the higher level Foursquare categories (Food, Nightlife, Travel \& Transport, Residences, Arts \& Entertainmens, Shops, College \& University
Outdoors \& Recreation, Professional \& Other places.)\footnote{\url{https://developer.foursquare.com/docs/resources/categories}}. 
Overall in the area covered by the North West Ambulance service, we have observed 
approximately $240$ thousand check-ins over the $11$ month period considered. We observe significant practical correlations between the Foursquare and ambulance call datasets despite the the fact that their time windows do not overlap (2011 versus 2013-2017).
% , acknowledging
% the possibility that a more temporally aligned source of check-in data could yield better results. 
% Twitter data invovled a real-time collection process for the Decahose (10\% of the total)~\footnote{\url{https://developer.twitter.com/en/docs/tweets/sample-realtime/overview/decahose
% }} for geolocated data. 
Finally, we have associated every Foursquare venue with an LSOA through a spatial join of the venues dataset with the polygons describing the boundaries of the LSOAs. 
% From ever place, we selected all tweets that had precise GPS coordinates  and fell within the bounding box defined by each of the LSOA areas we use in this work. 
% We consider online social media sources as general indicators of activity in areas. Despite the population penetration and usage biases, these sources offer the advantage of being able to track moving populations in almost real time and thus correspond to a promising
% source for predicting temporal trends of ambulance demand. 
\subsection{Spatial distribution of calls}
\begin{figure}[t]
\centering
\includegraphics[scale=0.20] {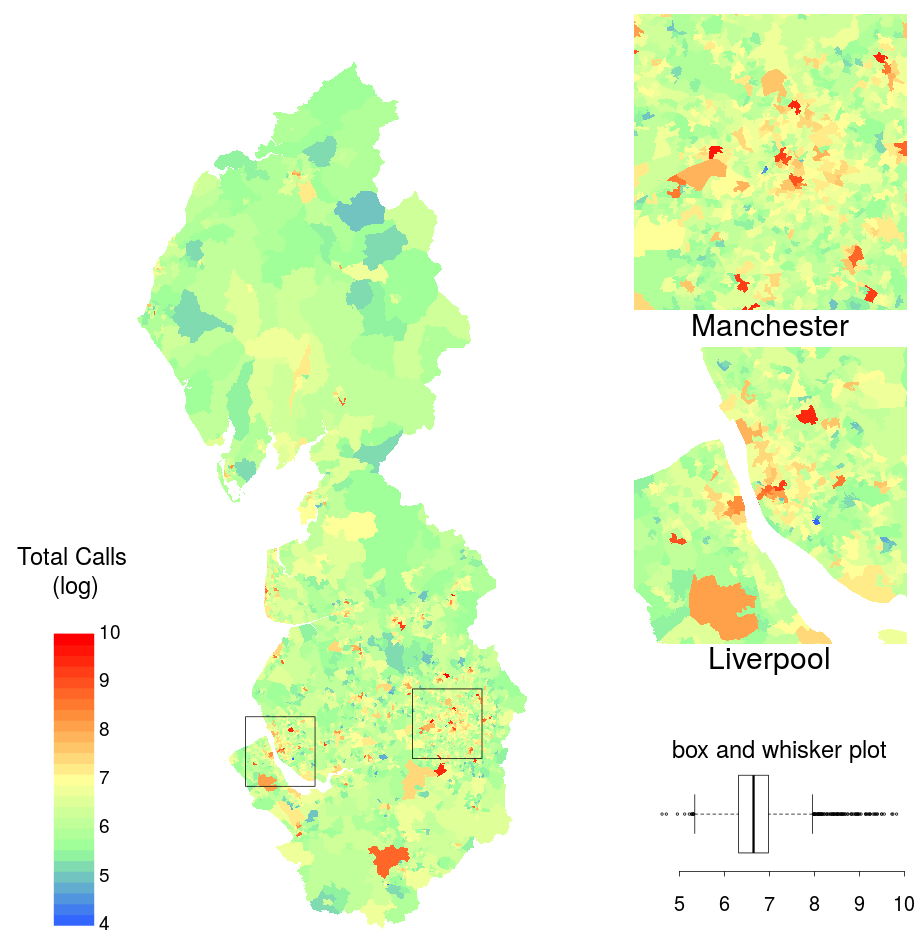}
\caption{Choropleth map for (natural log) for all calls with location of hospitals circled with zoomed  Manchester and Liverpool areas. Bottom right shows a box and whisker plot of the data. }
\label{totalcalls}
\end{figure}
The main focus of the present work is to predict the geographic variations of ambulance calls and the identification of the factors that drive their increase. To highlight the relevance of the question, we provide views on how ambulance calls are skewed across different geographic regions (LSOAs) in the North West of England.   The choropleth map for the total number of calls across the whole four year period is shown in Figure~\ref{totalcalls}, using the natural logarithm to represent the number of calls in each LSOA. 
One can see that the sizes of the LSOAs vary considerably to maintain approximately similar population sizes, with urban centres corresponding to smaller
geographic areas of higher population density. The squares on the main map show areas zoomed in to reveal more detail in the two main urban centers of the region, Manchester and Liverpool.
As can be observed, there is a significant variation in calls across areas.
% We set our goal of investigation next to be the prediction of such variations in calls together with the factors that may drive call activity. 
Table~\ref{toptenlsoas} shows the ten of those LSOAs with the highest numbers of calls. The highest rated is the LSOA with Manchester Airport, followed by a number of urban areas that are known to concentrate high commercial activity.
\begin{table}[]
\centering
\begin{tabular}{llll}
\hline
LSOA code & LSOA name                      & calls            & x mean \\
\hline
E01005316 & Manchester 053D                & 13274            & 14.7     \\
E01033658 & Manchester 054C                & 10216            & 11.3     \\
E01033653 & Manchester 055B                & 8401 & 9.3      \\
E01018326 & Cheshire Wes 034A & 7846 & 8.7      \\
E01012681 & Blackpool 006A                 & 7797             & 8.6      \\
E01005948 & Tameside 013A                  & 7397             & 8.2      \\
E01033760 & Liverpool 060C                 & 7095             & 7.9      \\
E01012736 & Blackpool 010D                 & 6833             & 7.6      \\
E01005758 & Stockport 014B                 & 6703             & 7.4      \\
E01033756 & Liverpool 061C                 & 6294             & 7.0       \\
\hline
\end{tabular}
\caption{Top 10 LSOAs for call volume over the 4 year period. Right column shows how many times greater than the mean of all LSOAs.}
\label{toptenlsoas}
\end{table}
~
\begin{table}[]
\centering
\begin{tabular}{lll}
\hline
Dispatch Code & Complaint                         & \%   \\
\hline
35            & Healthcare Practitioner Referral & 16.2 \\
17            & Falls                             & 13.3 \\
6             & Breathing Problems                & 10.8 \\
10            & Chest Pain                        & 9.1  \\
31            & Unconscious/Fainting              & 7.6  \\
26            & Sick Person                       & 7.1  \\
12            & Convulsions/Seizures              & 4.7  \\
25            & Psychiatric/Suicide               & 4.0  \\
23            & Overdose/Poisoning                & 3.0  \\
30            & Traumatic Injuries                & 2.5  \\
\hline
\end{tabular}
\caption{Ten most frequent dispatch codes ranked by frequency.}
\label{discodefreq}
\end{table}

\section{Analysis of Spatial Epidemiological Patterns}
\label{sec:analysis}
\begin{figure}%[b]{0.4\textwidth}
\centering
\includegraphics[scale=0.20] {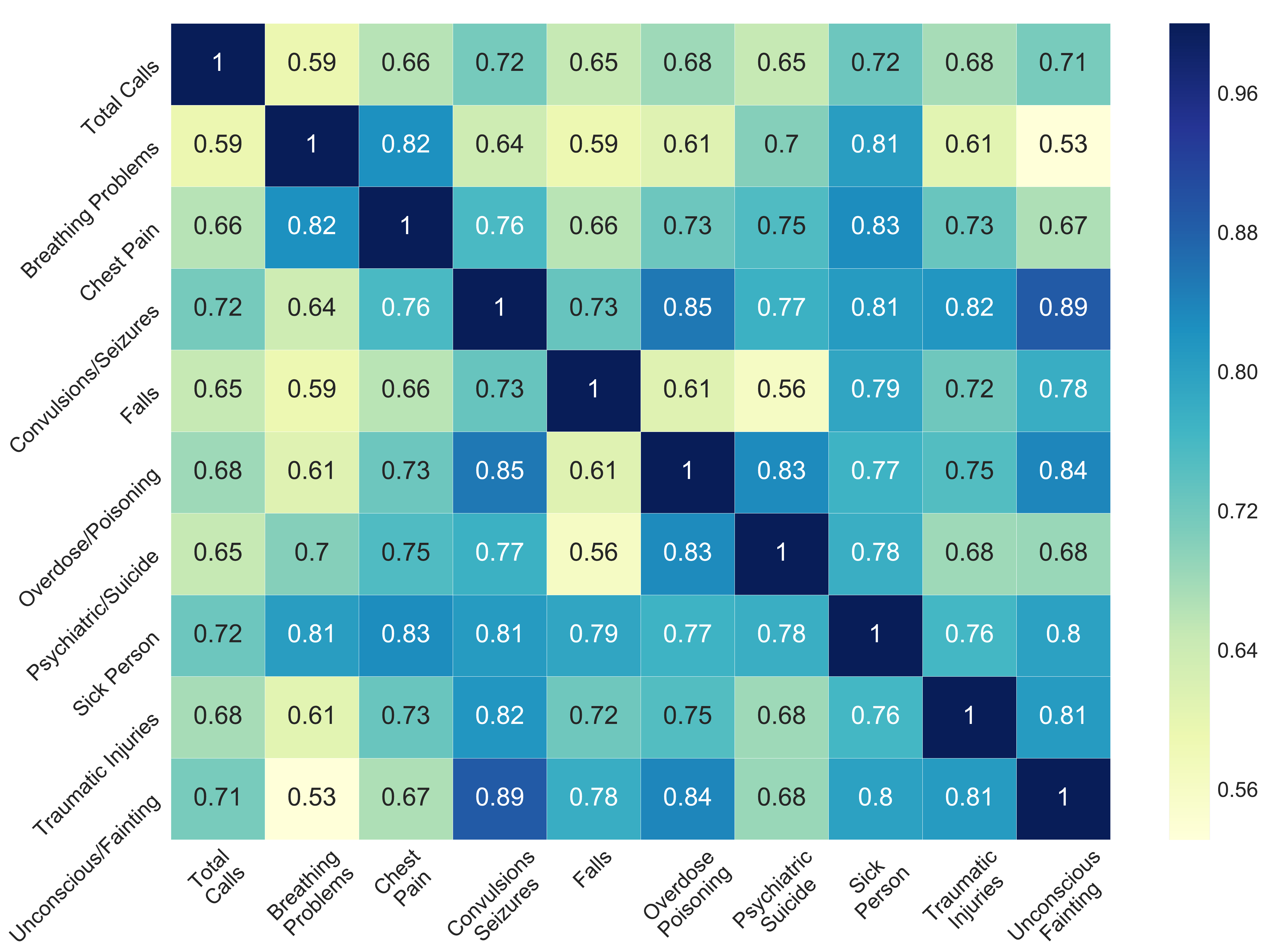}
\caption{Pearson correlation scores between variables representing pairs of health incidents.}
\label{incmatrix}
\end{figure}
In this section, we investigate to what extent socio-economic and demographic factors, together
with population related variables and data from location-based services (Foursquare) are associated
with different types of ambulance incidents. We perform a correlation-driven analysis reporting 
pearson correlation scores between pairs of variables across the various information layers. 
We have applied the Bonferroni correction when measuring the statistical significance of all correlation measurements and they were all significant
for the corresponding thresholds. 
\begin{figure}
    \centering
    \begin{subfigure}%[b]{0.4\textwidth}
	\centering
	\includegraphics[scale=0.05] {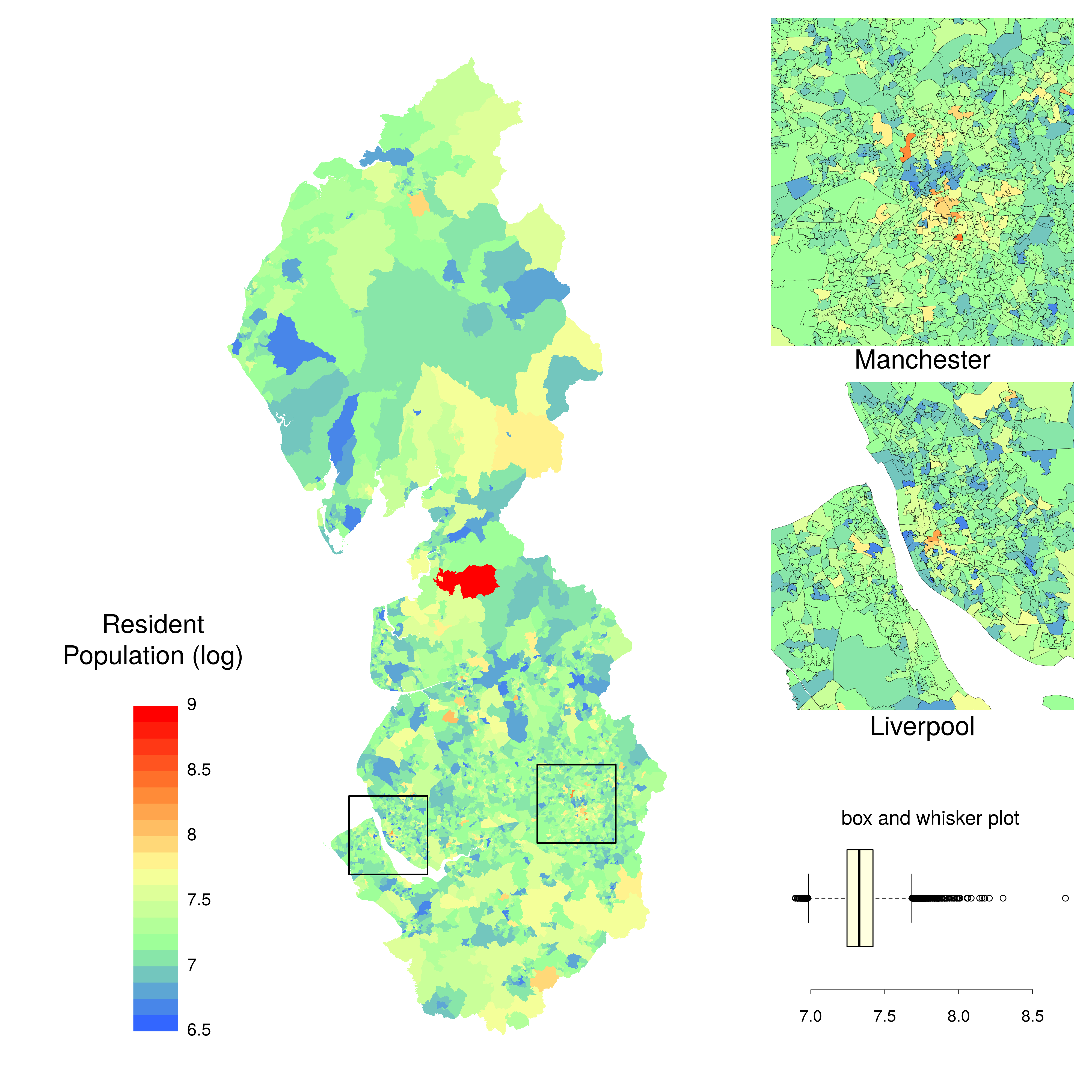}
	% \caption{Resident population}
	\label{respop}
    \end{subfigure}
~
    \begin{subfigure}%[b]{0.4\textwidth}
	\centering
	\includegraphics[scale=0.05] {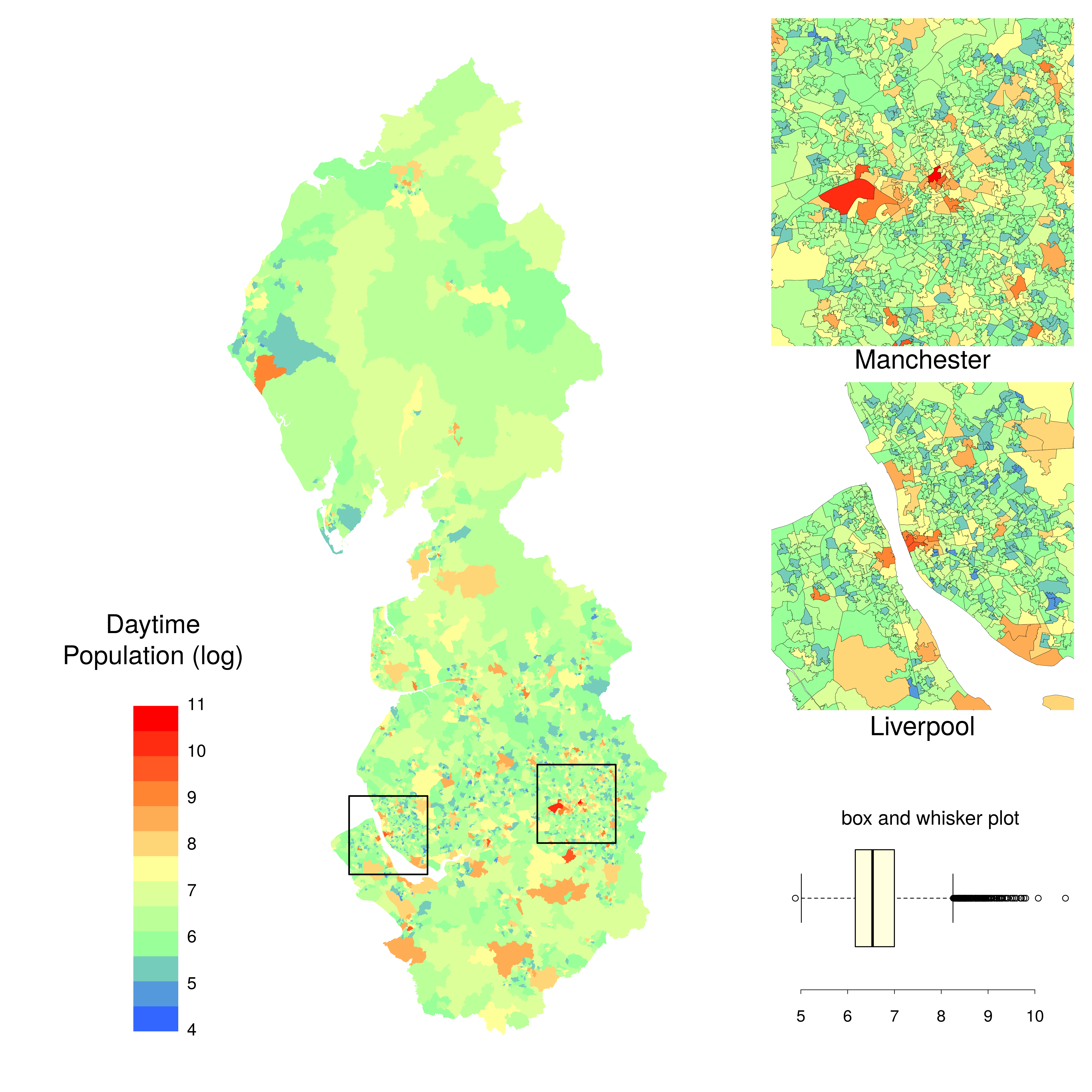}
	% \caption{Daytime Population... }
	\label{daytimepop}
    \end{subfigure}
    \caption{(a) Choropleth map and box-and-whisker plot of (log of) resident population size. (b) Choropleth map and box-and-whisker plot of (log of) daytime population size.}
    \label{popdistributions}
\end{figure}
\subsection{Incident frequencies and associations}
The ten most frequent dispatch code call for each age group resulted in a combined list of ten dispatch codes as given in Table~\ref{discodefreq}. They accounted for $78.3\%$ of all calls. Healthcare practitioner referral (Dispatch Code 35) is the most frequent and corresponds to calls made by doctors at healthcare facilities to transfer a patient to a hospital. 
In the analysis that follows next, we have filtered out calls that correspond to this incident code from the data as they relate primarily to 
operations within the health service (e.g. transfering patients to a hospital with more beds) and not to epidemiological traces we are interested in in the present work. 

Figure~\ref{incmatrix} shows the pearson's r correlation scores between the frequencies of all calls (Total Calls) per LSOA and the most frequent incident types (dispatch codes). 
A higher correlation score between two incident types implies a higher chance of them co-occuring spatially. 
\textit{breathing problems} and \textit{chest pain} correspond to one 
of the most related pairs (pearson's $r= 0.82$ ). As suggested by a correlation score of $r>0.83$, \textit{overdose/poisoning} related incidents are more likely to associate with the occurrence of \textit{psychiatric/suicide}, \textit{convulsion/seizuers} and \textit{unconscioous/fainting}  cases.
\textit{breathing problems} are more associated with \textit{chest pain} incidents and \textit{sick person} cases. These figures already indicate that different geographic areas 
may yield different epidemiological incidents in nature and hence the question becomes whether it is possible to identify the characteristics that makeup those areas and are contributing to these patterns. 
\subsection{Residential versus daytime populations}
\label{sec:population}
\begin{figure}
\centering
\includegraphics[scale=0.20] {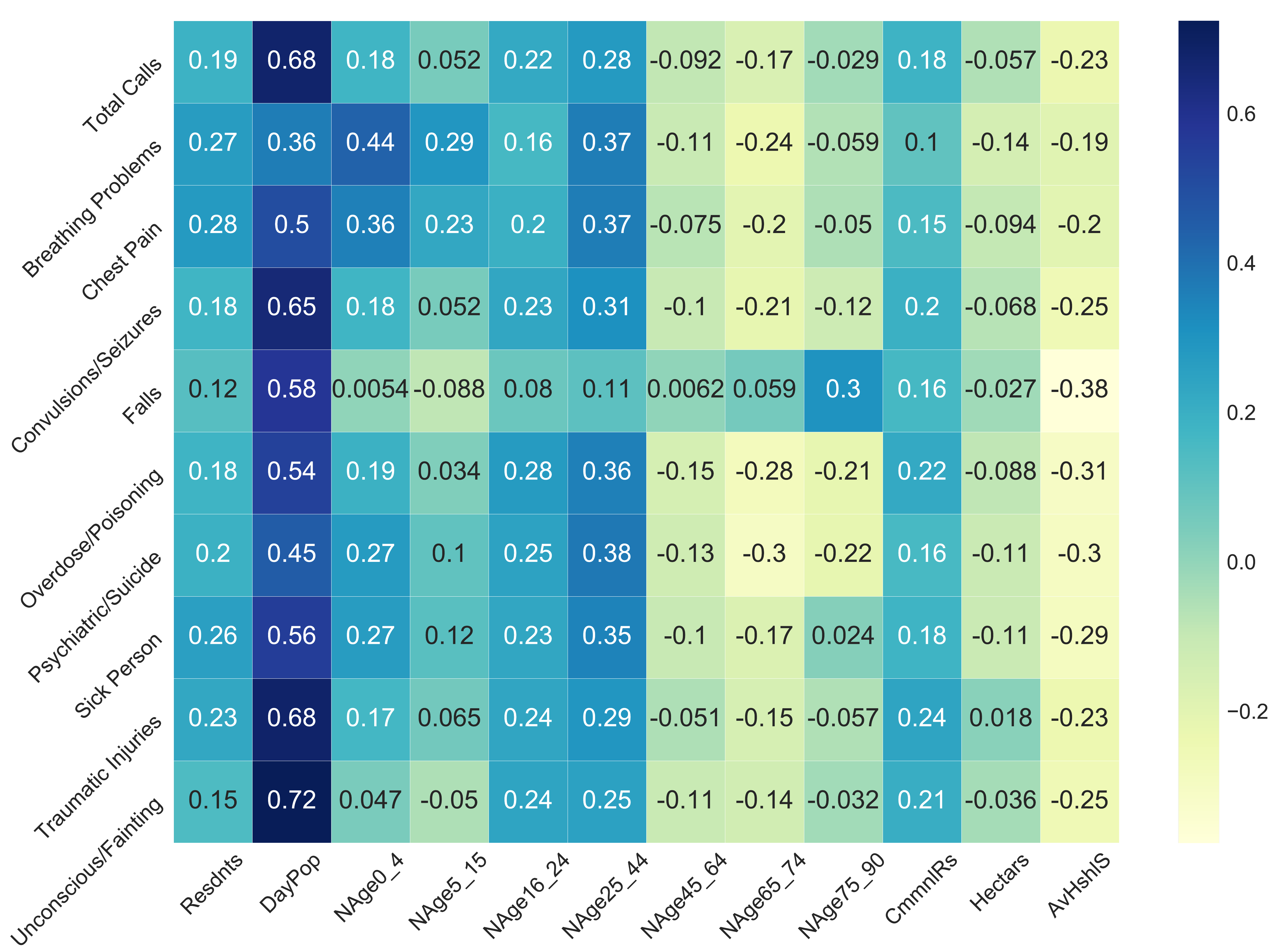}
\caption{Correlation matrix reporting pearson's r correlation scores between variables describing medical incidents attended by ambulances and a number of population, demographic and geographic variables.}
\label{corrmatrixDemos}
\end{figure}

In answering the question \textit{what geographic and demographic features influence the number of calls to an ambulance service?}, it seems reasonable to expect that, to some degree, the size of the population would be relevant. We present two population-based maps here.
While LSOAs were created with a goal of their being equal in population size (and therefore, varying in area), there was some considerable variation as Figure~\ref{popdistributions} shows. In fact, population size ranged from 988 (an urban area in Southport) to 6137 (a largely rural area south of Lancaster but which contains the university campus of Lancaster University), with a median of 1520. The city centres of Manchester and Liverpool show one or two higher than average populations, but in general green colours which correspond to lower population areas dominate.

Populations fluctuate constantly as people move about their daily activities. A characteristic example of such process is commuting, where typically 
large populations move from the more peripheral areas towards urban centers. In the present context, we can hypothesize that ambulance calls are likely to happen in  the areas where people are active and not simply in the areas where their residence is registered. With this in mind we have designed a new variable, namely \textit{daytime population} which is the sum of the \textit{workplace population} (described in the previous section) with residents younger than 16 and older than 74 year old. This aims to provide a proxy to the number of people active at a geographic area during working hours. 
Figure~\ref{popdistributions} shows the distribution of this feature geographically. 
%%%%%%%%%
Note that the colour scale is different from the previous plot showing residential population levels, with the range of daytime population being rather larger; from 132 in West Cumbria to 42357 in the City Centre of Manchester. Some areas of relatively low resident population have a much higher working population, with the reverse also being a possibility. Daytime and residential populations capture different temporal instances of population whereabouts and as we demonstrate in the following paragraphs are both important to estimating calls to the ambulance service. 
\begin{figure}
\centering
\includegraphics[scale=0.20] {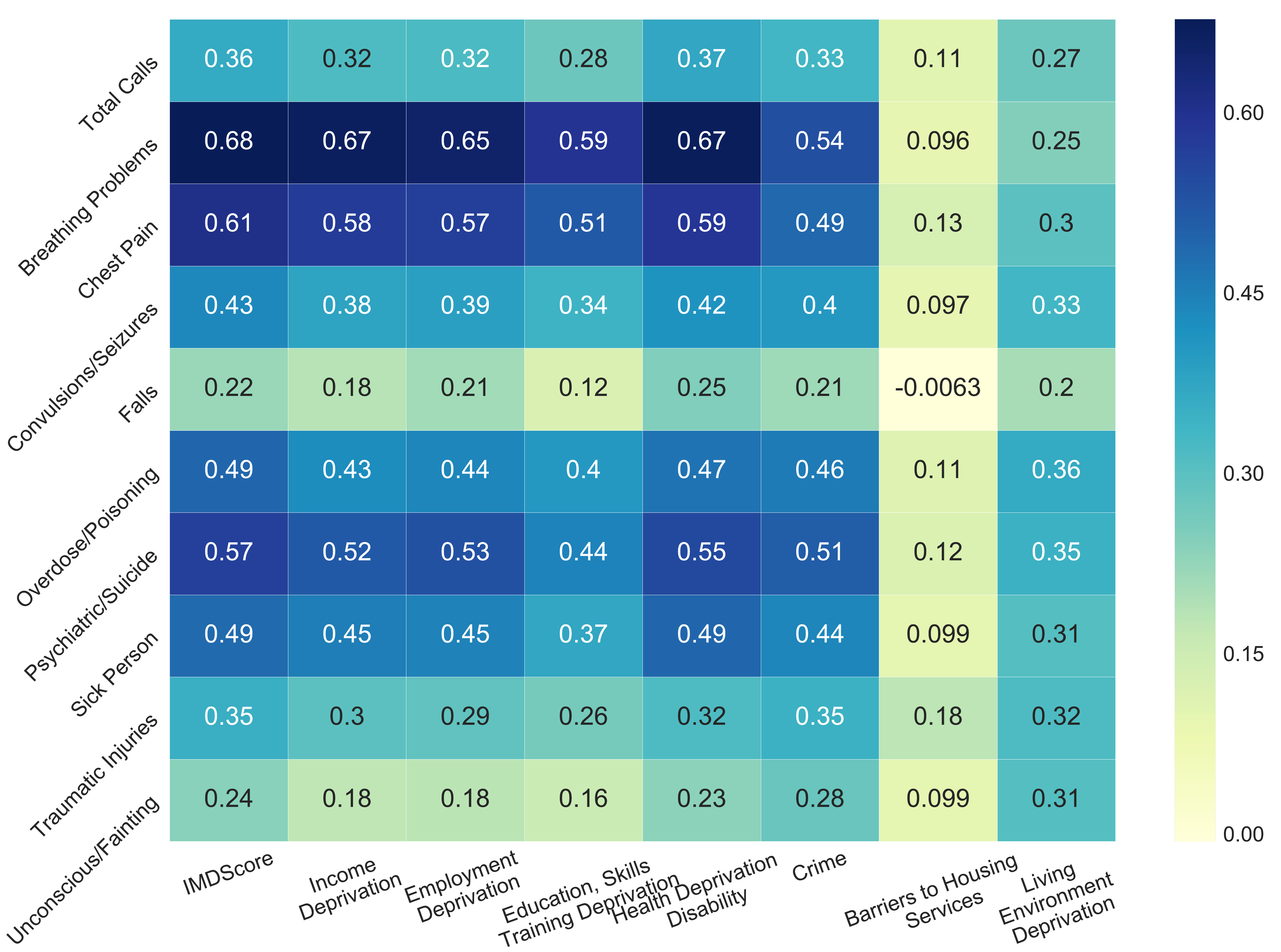}
\caption{Correlation matrix reporting pearson's r correlation scores between variables describing medical incidents attended by ambulances and a number of socio-economic indicators.}
\label{corrmatrixIMD}
\end{figure}
The daytime population levels appears to be a much more important indicator 
to residential population levels  as noted by the corresponding pearson's r with the total number of calls ($r=0.68$ for daytime population versus $0.18$ for residential). Daytime population
is strongly associated with the frequency of \textit{unconscious/fainting} incidents ($r=0.72$),
which are followed closely by \textit{traumatic injuries} ($r=0.68$), \textit{convulsions/seizures} $r=0.65$) and \textit{falls} ($r=0.58$). In terms of how different age group resident numbers 
explain ambulance call variations, the age group 25-44 sticks out as the one that contributes
consistently across different incident types (for most cases we have an $r>0.3$), with young 
kids of age 0-4 being associated to breathing problem related incidents ($r=0.44$) and the  
number of older people being the group mostly associated with \textit{falls} (pearson's $r=0.3$), in line with previous studies that have associated this age group with higher fall risk due 
to the presence of specific risk factors (e.g. weakness, unsteady gait, confusion and certain medications)~\cite{rubenstein2006falls}. 

\subsection{Socio-economic indicators}
\label{sec:socio}
We now investigate the relationship, between the index of multiple deprivation (IMD), the seven constituent metrics that it is comprised of and the frequencies and types of calls per LSOA. The pearson correlation scores between the 
different pairs of variables is shown in Figure~\ref{corrmatrixIMD}. 
The Index of Multiple Deprivation (IMD) scores a pearson's $r$ of $0.68$ with respect to \textit{breathing problems}  which is the most highly correlated incident type, 
followed by \textit{chest pain} incidents $r=0.61$ and \textit{psychiatric/suicide} incidents ($r=0.57$). While the overall index provides a general notion
of the deprivation levels of an area, its constituent metrics can shed light on the more 
specific factors that relate to ambulance calls. Income, employment, health and crime deprivation 
correlate highly with the incidents types noted above. Note that interestingly, the IMD score of an area does not yield an as high correlation score when considering other types of incidents such as \textit{falls}, \textit{traumatic injuries} or \textit{fainting}. These results highlight how population in deprived areas are essentially more likely to suffer a serious medical condition, perhaps due to lack of access to preemptive care and lifestyle related factors.
Links between deprivation of living standards have been identified before in the literature~\cite{bernard_emergency_1998,mccartney_how_2013,smith_relative_1992} with reported correlation score values in the range of $0.4-0.5$ across large geographic regions at national scale. Here we demonstrate that the link between deprivation and population health not only persists in smaller geographic scales, but in this case the geographic divide that exists amongst regions becomes larger.
\begin{figure}
    \centering
    \begin{subfigure}%[b]{0.4\textwidth}
	\centering
	\includegraphics[scale=0.33] {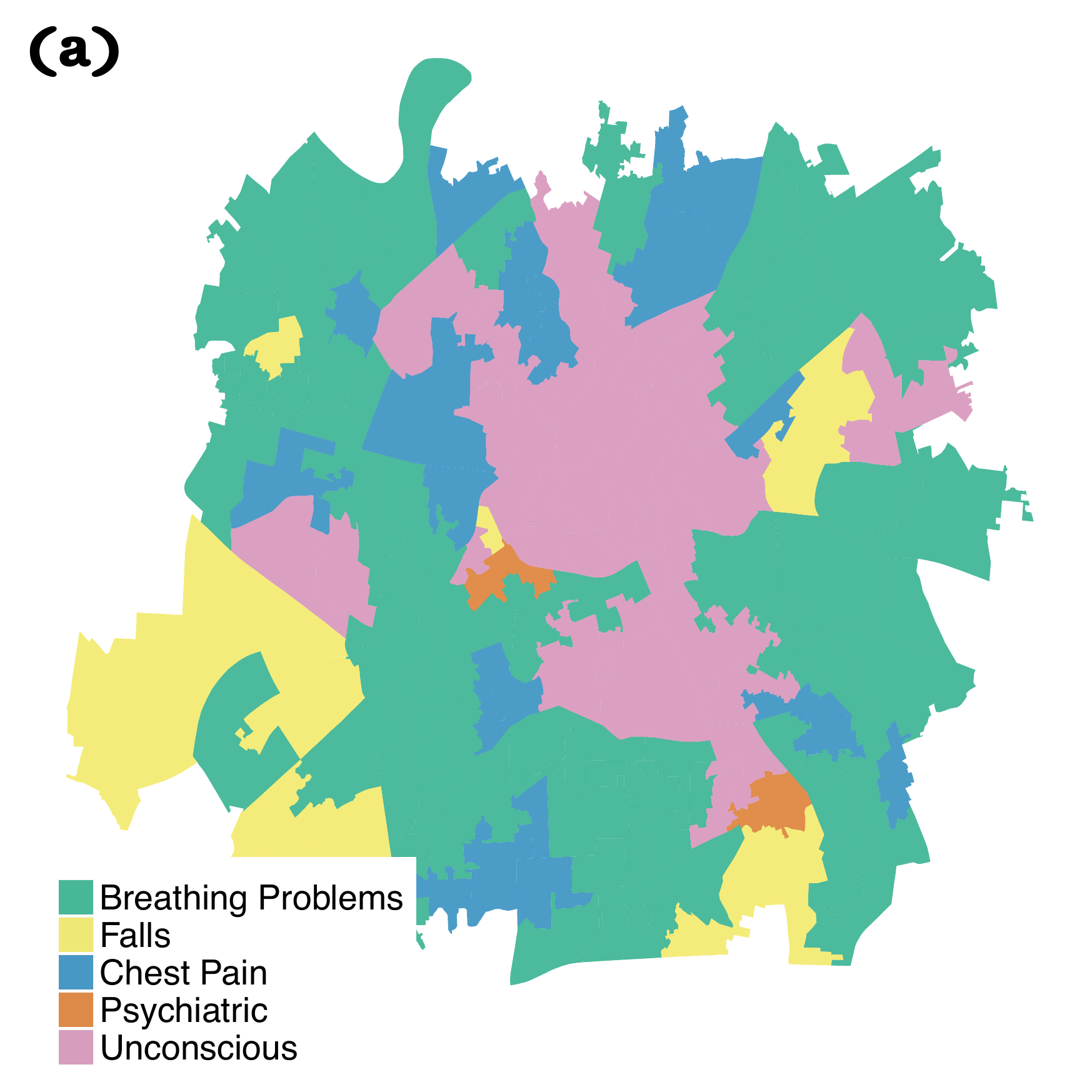}
	\label{maninc}
	% \caption{(a) Manchester medical incident types.}
    \end{subfigure}
~
    \begin{subfigure}%[b]{0.4\textwidth}
	\centering
	\includegraphics[scale=0.33] {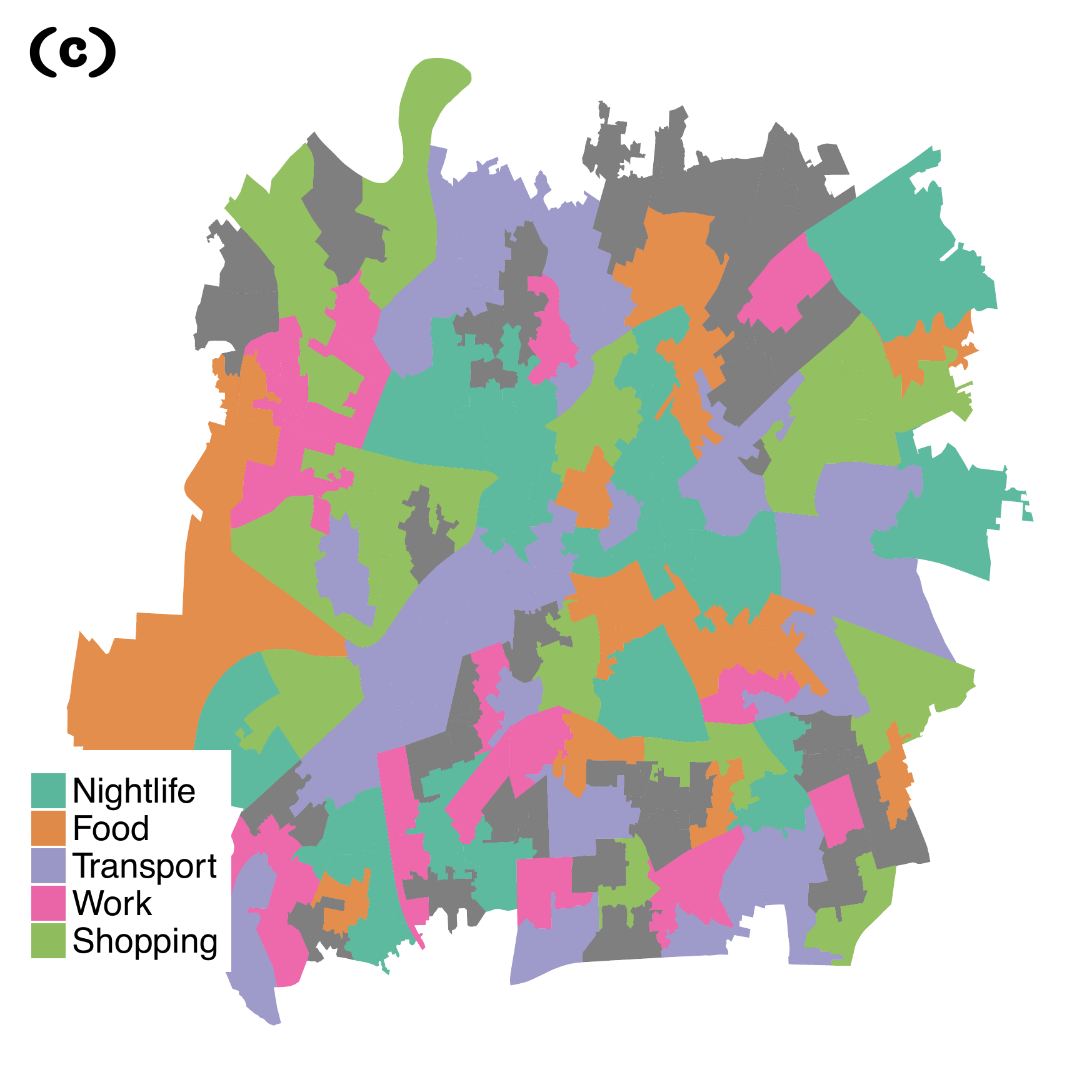}
	\label{mancheck}	
	% \caption{(c)}
    \end{subfigure}  
%     \centering
%     \begin{subfigure}%[b]{0.4\textwidth}
% 	\centering
% 	\includegraphics[scale=0.35] {FiguresTables/liv_incident_cat.pdf}
% 	\label{liveinc}
% 	% \caption{(a)}
%     \end{subfigure}
% \\
%     \begin{subfigure}%[b]{0.4\textwidth}
% 	\centering
% 	\includegraphics[scale=0.35] {FiguresTables/man_checkin_cat.pdf}
% 	\label{mancheck}	
% 	% \caption{(c)}
%     \end{subfigure}
% ~
%     \begin{subfigure}%[b]{0.4\textwidth}
% 	\centering
% 	\includegraphics[scale=0.35] {FiguresTables/liv_checkin_cat.pdf}
% 	\label{livcheck}
	% \caption{(d)}
    % \end{subfigure}
    \caption{Most common types of health incidents in central Manchester
    visually compared with the most popular categories in terms of number of check-ins in 
    the same area. Areas with no check-in activity are colored in grey.}
    \label{incidentsVsCheckins}
\end{figure}
\subsection{Digital traces of human mobility}
With the goal of understanding to what extent mobile web proxies of human urban activity 
such as Foursquare could capture geographic variations in ambulance calls
we visualise the two sources of data in Manchester in Figure~\ref{incidentsVsCheckins}. For a set of LSOAs in the center of the city, we plot the predominant categories of calls and check-ins respectively. Distinctive geographical patterns of the types of incidents can be observed for both cities, with \emph{unconscious/fainting} being the predominant epidemiological trend in the dense urban cores, whereas other types of emergencies are more characteristic of the peripheral areas. With comparison to human activity as derived from Foursquare check-ins, we can also note some activities which are more typical for the core as opposed to the periphery of the city such as \emph{nightlife} and \emph{shopping}. 

\begin{figure}
\centering
\includegraphics[scale=0.20] {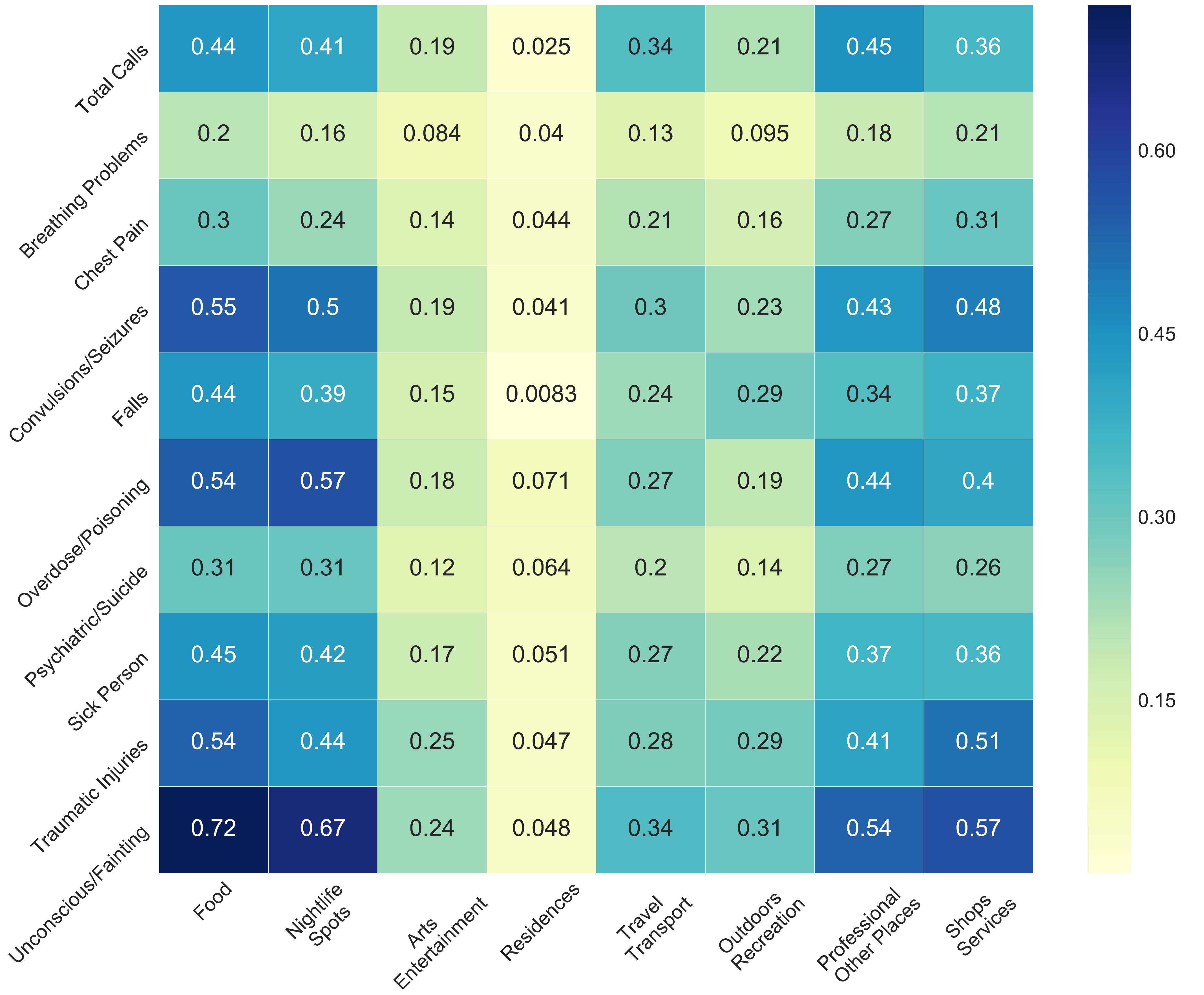}
\caption{Correlation matrix reporting Pearson's r scores between Foursquare categories and frequent types of medical
incidents.}
\label{fsqmatrix}
\end{figure}

We further quantify these relationships in Figure~\ref{fsqmatrix}, where a correlation matrix of call types and check-in types is presented for the whole North West region whereas the ambulance service operates.
We notice that specific types of human activity tend to be associated with particular calls on the small-scale of LSOAs. For example, the most common type of activity associated with the total number of calls is \emph{professional} places.  \emph{breathing problems} tend to be called in areas with a high number of \emph{shopping} and \emph{food} check-ins, similar to \emph{chest pain}. Cases of \emph{convulsions/seizures} are mostly associated with \emph{food, nightlife, shops} and work environments, as well as cases of \emph{falls, overdose/poisoning, traumatic injuries} and reports of \emph{sick person}. To a lesser extent, travel and outdoor activities were also related to such cases. However, the highest correlations found were between the \emph{food} and \emph{nightlife} categories and calls related to loss of consciousness (Pearson's $r=0.72$ and $r=0.67$ respectively). Overall, most types activities correlate with the \emph{unconscious} category of calls to a varying extent which was most expressed in dense urban centers where most calls are made (see Figure~\ref{totalcalls}). 
~

\section{Predicting Ambulance Calls}
\label{sec:results}
\begin{figure}[t]
\centering
\includegraphics[scale=0.35] {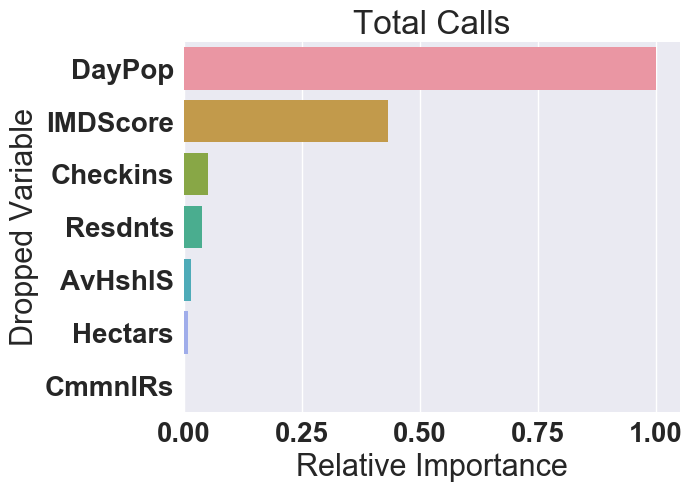}
\caption{Relative importance of variables to the prediction of the total number of calls.}
\label{relImpAll}
\end{figure}
\begin{table*}[]
\begin{tabular}{lcccc}
\textbf{All Calls}:&$Adj R^{2}=0.702$ \\ % &&& \textbf{Breathing Problems}:&$Adj R^{2}=0.645$ \\
\hline
                   & \textbf{coef} & \textbf{std err} & \textbf{t} & \textbf{P$>$$|$t$|$}  \\
\midrule
\textbf{IMDScore}  &       0.0784  &        0.003     &    24.563  &         0.000             \\
\textbf{Checkins}  &       0.2369  &        0.025     &     9.547  &         0.000           \\
\textbf{DayPop}    &       0.8359  &        0.022     &    37.676  &         0.000            \\
\textbf{Resdnts}   &       0.0989  &        0.013     &     7.514  &         0.000           \\
\textbf{CmmnlRs}   &       0.0311  &        0.023     &     1.347  &         0.178       \\
\textbf{AvHshlS}   &      -0.0413  &        0.010     &    -4.199  &         0.000          \\
\textbf{Hectars}   &      -0.0392  &        0.011     &    -3.647  &         0.000            \\
% \textbf{intercept} &   -1.267e-16  &        0.001     &  -1.9e-13  &         1.000         \\
\bottomrule
\end{tabular}
~
%breathing problems
\begin{tabular}{lcccc}
\textbf{Breathing \newline~problems}:& $Adj R^{2}=0.645$ \\
\hline
                   & \textbf{coef} & \textbf{std err} & \textbf{t} & \textbf{P$>$$|$t$|$}  \\
\midrule
\textbf{IMDScore}  &       0.3077  &        0.005     &    57.846  &         0.000         \\
\textbf{Checkins}  &      -0.0095  &        0.042     &    -0.229  &         0.819         \\
\textbf{DayPop}    &       0.7904  &        0.037     &    21.292  &         0.000          \\
\textbf{Resdnts}   &       0.4577  &        0.022     &    20.920  &         0.000          \\
\textbf{CmmnlRs}   &      -0.2449  &        0.039     &    -6.339  &         0.000           \\
\textbf{AvHshlS}   &      -0.1286  &        0.016     &    -7.858  &         0.000           \\
\textbf{Hectars}   &      -0.1677  &        0.018     &    -9.433  &         0.000            \\
% \textbf{intercept} &   -6.256e-17  &        0.001     & -5.66e-14  &         1.000           \\
\bottomrule
\end{tabular}
~
\newline
%chest pain
\begin{tabular}{lcccc}
\textbf{Chest Pain}:& $Adj R^{2}=0.645$ \\
\hline
                   & \textbf{coef} & \textbf{std err} & \textbf{t} & \textbf{P$>$$|$t$|$}   \\
\midrule
\textbf{IMDScore}  &       0.2096  &        0.004     &    49.674  &         0.000            \\
\textbf{Checkins}  &       0.1346  &        0.033     &     4.084  &         0.000             \\
\textbf{DayPop}    &       0.8262  &        0.029     &    28.055  &         0.000             \\
\textbf{Resdnts}   &       0.3181  &        0.017     &    18.328  &         0.000             \\
\textbf{CmmnlRs}   &      -0.1038  &        0.031     &    -3.386  &         0.001            \\
\textbf{AvHshlS}   &      -0.0779  &        0.013     &    -6.002  &         0.000           \\
\textbf{Hectars}   &      -0.0764  &        0.014     &    -5.414  &         0.000           \\
\bottomrule
\end{tabular}
~
%convulsions seizures
\begin{tabular}{lcccc}
\textbf{Convulsions}:& $Adj R^{2}=0.645$ \\
\hline
                   & \textbf{coef} & \textbf{std err} & \textbf{t} & \textbf{P$>$$|$t$|$}   \\
\midrule
\textbf{IMDScore}  &       0.0958  &        0.003     &    33.775  &         0.000            \\
\textbf{Checkins}  &       0.4229  &        0.022     &    19.101  &         0.000            \\
\textbf{DayPop}    &       0.6123  &        0.020     &    30.940  &         0.000           \\
\textbf{Resdnts}   &       0.0874  &        0.012     &     7.495  &         0.000             \\
\textbf{CmmnlRs}   &       0.0443  &        0.021     &     2.151  &         0.032             \\
\textbf{AvHshlS}   &      -0.0560  &        0.008     &    -6.929  &         0.000             \\
\textbf{Hectars}   &      -0.0378  &        0.009     &    -3.986  &         0.000          \\
\bottomrule
\end{tabular}
~
\newline
%falls 
\begin{tabular}{lcccc}
\textbf{Falls}:& $Adj R^{2}=0.448$ \\
\hline
                   & \textbf{coef} & \textbf{std err} & \textbf{t} & \textbf{P$>$$|$t$|$}  \\
\midrule
\textbf{IMDScore}  &       0.0358  &        0.004     &     9.335  &         0.000           \\
\textbf{Checkins}  &       0.2065  &        0.030     &     6.901  &         0.000           \\
\textbf{DayPop}    &       0.6573  &        0.027     &    24.576  &         0.000            \\
\textbf{Resdnts}   &       0.1099  &        0.016     &     6.970  &         0.000             \\
\textbf{CmmnlRs}   &      -0.0283  &        0.028     &    -1.018  &         0.309            \\
\textbf{AvHshlS}   &      -0.2200  &        0.012     &   -18.658  &         0.000             \\
\textbf{Hectars}   &      -0.0278  &        0.013     &    -2.170  &         0.030          \\
\bottomrule
\end{tabular}
~
%overdose poisoning 
\begin{tabular}{lcccc}
\textbf{Overdose}:& $Adj R^{2}=0.616$ \\
\hline
                   & \textbf{coef} & \textbf{std err} & \textbf{t} & \textbf{P$>$$|$t$|$}   \\
\midrule
\textbf{IMDScore}  &       0.0867  &        0.002     &    36.193  &         0.000           \\
\textbf{Checkins}  &       0.4574  &        0.019     &    24.476  &         0.000            \\
\textbf{DayPop}    &       0.2700  &        0.017     &    16.166  &         0.000            \\
\textbf{Resdnts}   &       0.0872  &        0.010     &     8.849  &         0.000           \\
\textbf{CmmnlRs}   &       0.0916  &        0.017     &     5.270  &         0.000            \\
\textbf{AvHshlS}   &      -0.0884  &        0.007     &   -11.999  &         0.000          \\
\textbf{Hectars}   &      -0.0409  &        0.008     &    -5.108  &         0.000           \\
\bottomrule
\end{tabular}
~
\newline
%psychiatric / suicide 
\begin{tabular}{lcccc}
\textbf{Psychiatric}:& $Adj R^{2}=0.553$ \\
\hline
                   & \textbf{coef} & \textbf{std err} & \textbf{t} & \textbf{P$>$$|$t$|$}   \\
\midrule
\textbf{IMDScore}  &       0.1766  &        0.004     &    40.795  &         0.000           \\
\textbf{Checkins}  &       0.2215  &        0.034     &     6.555  &         0.000             \\
\textbf{DayPop}    &       0.5762  &        0.030     &    19.083  &         0.000            \\
\textbf{Resdnts}   &       0.2015  &        0.018     &    11.309  &         0.000           \\
\textbf{CmmnlRs}   &       0.0103  &        0.031     &     0.328  &         0.743        \\
\textbf{AvHshlS}   &      -0.1679  &        0.013     &   -12.604  &         0.000           \\
\textbf{Hectars}   &      -0.0923  &        0.014     &    -6.385  &         0.000          \\
\bottomrule
\end{tabular}
~
%sick person
\begin{tabular}{lcccc}
\textbf{Sick person}:& $Adj R^{2}=0.609$ \\
\hline
                   & \textbf{coef} & \textbf{std err} & \textbf{t} & \textbf{P$>$$|$t$|$}  \\
\midrule
\textbf{IMDScore}  &       0.1414  &        0.004     &    35.303  &         0.000           \\
\textbf{Checkins}  &       0.4558  &        0.031     &    14.579  &         0.000          \\
\textbf{DayPop}    &       0.6434  &        0.028     &    23.028  &         0.000         \\
\textbf{Resdnts}   &       0.2833  &        0.016     &    17.207  &         0.000       \\
\textbf{CmmnlRs}   &      -0.0639  &        0.029     &    -2.198  &         0.028           \\
\textbf{AvHshlS}   &      -0.1747  &        0.012     &   -14.186  &         0.000           \\
\textbf{Hectars}   &      -0.1063  &        0.013     &    -7.947  &         0.000            \\
\bottomrule
\end{tabular}
~
\newline
%traumatic injuries
\begin{tabular}{lcccc}
\textbf{Traumatic injuries}:& $Adj R^{2}=0.618$ \\
\hline
                   & \textbf{coef} & \textbf{std err} & \textbf{t} & \textbf{P$>$$|$t$|$}  \\
\midrule
\textbf{IMDScore}  &       0.0914  &        0.004     &    25.758  &         0.000            \\
\textbf{Checkins}  &       0.4481  &        0.028     &    16.175  &         0.000            \\
\textbf{DayPop}    &       0.8070  &        0.025     &    32.593  &         0.000            \\
\textbf{Resdnts}   &       0.1253  &        0.015     &     8.582  &         0.000             \\
\textbf{CmmnlRs}   &       0.1084  &        0.026     &     4.210  &         0.000           \\
\textbf{AvHshlS}   &      -0.0607  &        0.011     &    -5.559  &         0.000            \\
\textbf{Hectars}   &       0.0335  &        0.012     &     2.825  &         0.005          \\
\bottomrule
\end{tabular}
~
%unconscious fainting
\begin{tabular}{lcccc}
\textbf{Unconscious}:& $Adj R^{2}=0.685$ \\
\hline
                   & \textbf{coef} & \textbf{std err} & \textbf{t} & \textbf{P$>$$|$t$|$}   \\
\midrule
\textbf{IMDScore}  &       0.0371  &        0.002     &    17.467  &         0.000            \\
\textbf{Checkins}  &       0.5552  &        0.017     &    33.448  &         0.000             \\
\textbf{DayPop}    &       0.4430  &        0.015     &    29.865  &         0.000             \\
\textbf{Resdnts}   &       0.0445  &        0.009     &     5.095  &         0.000           \\
\textbf{CmmnlRs}   &       0.0372  &        0.015     &     2.409  &         0.016           \\
\textbf{AvHshlS}   &      -0.0496  &        0.007     &    -7.590  &         0.000         \\
\textbf{Hectars}   &      -0.0211  &        0.007     &    -2.972  &         0.003          \\
\bottomrule
\end{tabular}
\caption{\label{linearResults} Summary of linear regression results.}
\end{table*}

Our aim next becomes to predict the number of ambulance calls for each location by considering a set of the variables discussed in the previous section in an ordinary
least squares (OLS) linear regression model. Formally, given an area $i$ and then the number of calls that originated from the area, $y_{i}$ is set to be approximated by the linear following
relationship:
\begin{equation}
y_{i} = \mathbf{x_i}^{T} \mathbf{\beta} + \epsilon_{i}
\end{equation}
where $x_i$ represents a predictor variable, 
and  where $\beta$ is a $p\times1$ vector of unknown parameters
where $p$ is the number of input variables. $\epsilon_{i}$
in this setting are unobserved scalar random variables (errors) which account for the discrepancy between the actually observed responses $y_i$ and the predicted outcomes. 
To alleviate colinearity effects we have not used variables that are inherently related (e.g. age group frequencies and residential population). All variables have been standardized 
by substracting the corresponding mean and dividing by the standard deviation.  
As a metric of assessment for the prediction task we use the adjusted $R^{2}$ which provides an indication of how much of the variance is explained by the model
compared to the total variance of variable $y$ taking into account the number of independent variables. Overall, we examine ten prediction tasks, one for the overall number of calls in an area, and one for each of the nine most popular incidents types.
\begin{figure*}[t]
\centering
\includegraphics[scale=0.22] {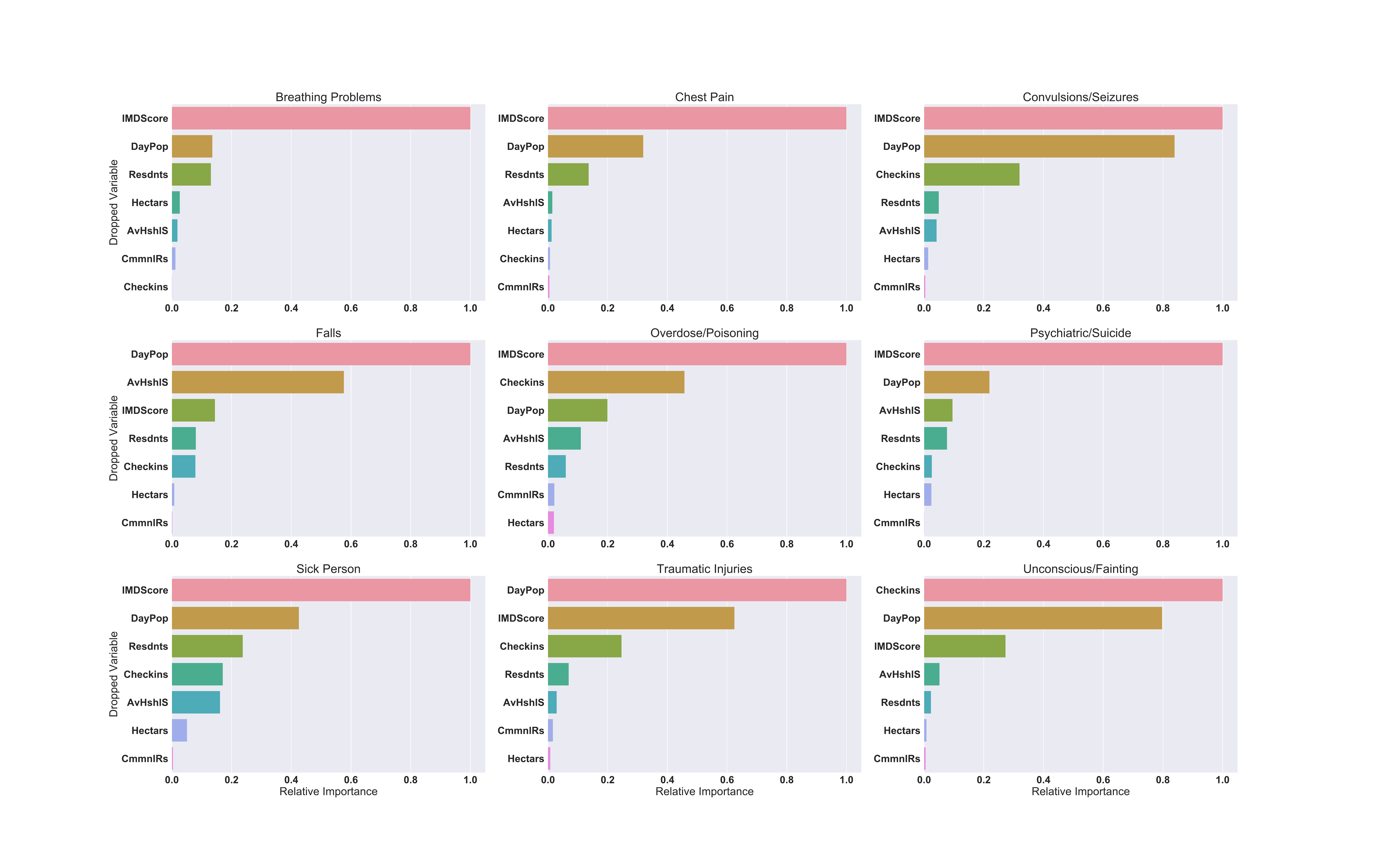}
\caption{Relative importance of variables to the prediction for different types incidents.}
\label{relImpTypes}
\end{figure*}
\paragraph{\textbf{Evaluation results}}  
In Table~\ref{linearResults} we present the coefficients of the variables of the linear regression models built for four out of ten prediction tasks considered. The adjusted $R^{2}$ values are considerably high in most cases with an $R^{2}=0.702$ 
being achieved when the total number of calls is considered as the dependent variable. The model attains an  $R^{2}=0.645$ for \textit{breathing problems} and for cases of \textit{chest pain} and  \textit{sick person} incidents the values of $R^{2}$ remained above $0.6$.  The lowest number was recorded for the cases of \textit{falls} with an $R^{2}=0.447$ whereas the rest of the incident types were predicted with values above $0.5$. In all cases, with the sole exception of breathing problems the Foursquare variable (check-ins) corresponded to a statistically significant case and in all such cases the sign of the variable was positive implying that a higher number of check-ins in  an area is in general associated with a higher number of calls. 

To assess the importance of the different information signals in explaining the variance of different types of incidents we run the following experiment. For each incident type, we removed each of the variables and measured the reduction in
$R^{2}$. To obtain then the \textit{relative importance} of a variable we simply measured the reduction in $R^{2}$ associated to it with respect to the maximum 
reduction attained by any of the variables. The barplot in Figure~\ref{relImpAll}
show the relative importance of each predictor for the total number of calls, whereas
in Figure~\ref{relImpTypes} we plot the results for all nine types of incidents. 
Notably, the index of multiple deprivation appears to be the most important variable
in the majority of cases with \textit{daytime} population explaining best \textit{falls} and \textit{traumatic injuries} cases. Foursquare check-ins correspond to the third most important variable when considering the total number of calls, whereas for 
incidents of \textit{unconscious/fainting} it becomes the most important indicator. An explanation for the performance of the Foursquare variable is the fact that most \textit{unconscious/fainting} incidents occur in the city centers and the service's usage patterns tend to be associated with activity in commercial, food and nightlife areas. 
Its importance for the \textit{overdose/poisoning} case of incidents where it scores higher even than \textit{daytime} population activity points in this direction.

\section{Conclusion}
Our results highlight the opportunity that arises from using 
data from online media sources and the mobile web to power the operation
of medical services critical for citizens. Limitations in using such 
data sources in the present context relate to biases in mobile application
usage patterns amongst others. Daytime population levels 
have been an important predictor of ambulance calls and a clear improvement
to simply using residential population information, yet it still represents
a static signal about the activity levels of an area. Populations
fluctuate constantly and so a promising future direction would be to exploit real time
digital datasets from location-based services to model medical incident activity 
not only across geographies, but also over time. The importance of deprivation
indicators in explaining geographic variations of ambulance calls provide 
an additional reminder of the large divides that exist in our society. Providing
evidence through data driven analysis of population activity and government
collected socio-economic indicators as we have done in the present work is 
an important step to bridge this gap by informing relevant policies.

\small
\bibliographystyle{plain}
\bibliography{biblio} 

\end{document}